\documentclass[twocolumn]{aastex62}

\usepackage{times,flushend,amsmath,amssymb}

\newcommand{\mc}[1]{\multicolumn{2}{c}{#1}}

\received{\today}
\revised{\today}
\accepted{\today}
\submitjournal{ApJ}

\shorttitle{The nature of the eccentric doubled-lined eclipsing  binary system KIC\,2306740}
\shortauthors{Ko{\c c}ak et al.}

\begin{document}

\title{The nature of the eccentric doubled-lined eclipsing  binary system KIC\,2306740 with {\it{Kepler}} space photometry}

\author[0000-0002-3000-9829]{D. Ko{\c c}ak}
\affil{Department of Astronomy and Space Sciences, University of Ege, 35100, Bornova--{\.I}zmir, Turkey}

\author[0000-0003-2380-9008]{K. Yakut}
\affiliation{Department of Astronomy and Space Sciences, University of Ege, 35100, Bornova--{\.I}zmir, Turkey}
\affiliation{Institute of Astronomy, University of Cambridge, Madingley Road, Cambridge CB3 0HA, UK}

\author{J. Southworth}
\affiliation{Astrophysics Group, Keele University, Staffordshire ST5 5BG, UK}

\author{P. P. Eggleton}
\affiliation{Lawrence Livermore National Laboratory, 7000 East Ave, Livermore, CA94551, USA}
 
\author[0000-0002-7874-7820]{T. {\.I}{\c c}li}
\affiliation{Department of Astronomy and Space Sciences, University of Ege, 35100, Bornova--{\.I}zmir, Turkey}
 
\author[0000-0002-1556-9449]{ C. A. Tout}
\affiliation{Institute of Astronomy, University of Cambridge, Madingley Road, Cambridge CB3 0HA, UK}

\author{ S.\ Bloemen}
\affiliation{Instituut voor Sterrenkunde, Katholieke Universiteit Leuven, Celestijnenlaan 200D, B-3001 Leuven, Belgium}
\affiliation{Department of Astrophysics, IMAPP, University of Nijmegen, PO Box 9010, 6500 GL Nijmegen,, the Netherlands}



\begin{abstract}

We present a detailed study of KIC\,2306740, an eccentric 
double-lined eclipsing binary system.
\textit{Kepler} satellite data were combined with spectroscopic data  
obtained with the 4.2\,m William Herschel Telescope (WHT).  
This allowed us to determine precise orbital and
physical parameters of this relatively long period ($P=10\fd 3$) and slightly
eccentric, ($e=0.3$)  binary system. 
The physical parameters have been determined as $M_1 = 1.194\pm0.008$\,M$_{\odot}$, $M_2 =
1.078\pm0.007$\,M$_{\odot}$, $R_1 = 1.682\pm0.004$\,R$_{\odot}$, $R_2
= 1.226\pm0.005$\,R$_{\odot}$, $L_1 = 2.8\pm0.4$\,L$_{\odot}$, $L_2
= 1.8\pm0.2$\,L$_{\odot}$ and orbital seperation $a = 26.20\pm0.04$\,R$_{\odot}$ through simultaneous solutions of \textit{Kepler} light curves and of the WHT radial velocity data.
Binarity effects were extracted from the light curve in order to study intrinsic variations in the residuals.  
Five significant and more than 100~combination frequencies were detected. 
We modeled the binary system assuming non-conservative evolution models with the Cambridge {\sc stars (twin)} code and we show evolutionary tracks of the components in the $\log L - \log T$ plane, the $\log R - \log M$ plane and the $\log P - \rm age$ plane for both spin and orbital periods together with eccentricity $e$ and $\log R_1$. 
The model of the non-conservative processes in the code led the system to evolve to the observed system parameters in roughly $5.1\,$ Gyr.

\end{abstract}



\section{Introduction} \label{sec:intro}

Double-lined eclipsing detached binary stars are an important source for accurately determining the physical parameters of the component stars \citep{2010A&ARv..18...67T}. Pulsations can be used for determining physical parameters as well as understanding stellar structure. Pulsating components in binary systems play an important role in understanding stellar structure because they are effectively laboratories for investigating stellar interiors \citep{2013EAS....64..323A}.Therefore, having a pulsating component in a binary provides an independed verification of stellar parameters.
Using continuous high precision observations from the  CoRoT, \textit{Kepler} and TESS satellites' data provide the opportunity  to study a variety of pulsating stars in binary systems.
Recently many observational results for different type pulsating
stars, including those of binary components, have been studied in the
literature \citep[e.g.][]{2004ApJ...604..800W, 2007A&A...467..647Y, 2011ApJS..197....4W, 2014A&A...563A..59M, 2018MNRAS.474.4322M, 2018MNRAS.475..478Q, 2019MNRAS.482.1231J}.
Pulsating stars in close binary systems have been
discussed in detail by many authors \citep[e.g.][]{1975A&A....41..329Z, Aerts04, 2005ASPC..333..138B, 2005ApJ...634..602R, 2010aste.book.....A, 2015ASSL..408..169H, 2020MNRAS.497L..19S}.

The {\it{Kepler}} satellite observed more than 200,000 stars, including some with planetary companions,
binary/multiple stellar systems, and pulsating stars, to obtain very high precision photometry \citep{{2010ApJ...713L..79K},  {2010Sci...327..977B}, {2010PASP..122..131G}, {2011AJ....142..112B}}.These photometric results have found previously unknown variations, providing further constraints to current models. The precision of the {\it{Kepler}} observations allow us to disentangle low-amplitude variation in a binary star system. One such system is KIC 2306740 which will be the focus of this work.

KIC~2306740  ($P=10\fd3$, $e=0.3$, V=$13\fm08$, $K_p=13\fm55$) is an eccentric double-lined detached eclipsing binary system that was discovered by \textit{Kepler} satellite.
Some basic parameters for the system given in Table~\ref{Table:inPar}.
The first preliminary binary solution of the system was found by \cite{prsa2011}. 
They refined the orbital period as $10\fd 30399\pm0\fd00003$ and temperature ratio ($T_2/T1$) as 0.834. \cite{Kjurkchieva2017} subsequently estimated the relative radii of the components as $r_1=0.0.612$ and $r_2=0.0600$, mass ratio as 0.972, the orbital eccentricity as 0.299, and the argument of periastron as $275^{\rm o}$. However, the parameters obtained in the current work are quite different
from those two studies. This is because of the careful interactive analysis of the light
curve (LC) and radial velocity (RV) data made in this study, rather than the automated LC modeling. 

In this paper we study the binary nature of the system as well its and rotational behaviour
using the \textit{Kepler} data combined with a set of high precision RVs.
KIC\,2306740 was observed by {\it{Kepler}} in quarters Q0 to Q16.
The new spectroscopic observations and data analysis of radial velocities are described in Section~2. We present the {\it{Kepler}} data and light curve solution of the system in Section~3. Using the radial velocity and light curve solution we obtained the physical parameters of the system in Section~4. Light variation outside eclipses is discussed in Section~5. Section~6 contains a discussion of the possible evolutionary state of the system and our conclusions.

\begin{table}\centering
	\caption{Basic parameters for KIC\,2306740. $B$ and~$V$ color values are taken from \citet{zacharias2005}
		and other parameters are taken from the Kepler Input Catalogue, Gaia and Simbad.}
	\begin{tabular}{lc}
		\hline
		Parameter               	      &    Value      \\
		\hline
		2MASS ID                	     & 19290475+3741535  \\
		Gaia ID              				& 2051885033280089216 \\
		$\alpha_{2000}$         	 & 19 29 04.75   \\
		$\delta_{2000}$         	  & +37 41 53.5   \\
		B                      				  & $13\fm 91$    \\
		V                      				  & $13\fm 08$    \\
		R                       			  & $12\fm 92$    \\
		G (Gaia)                        & $13\fm 473$    \\
		J (2MASS)              			& $12\fm 297$    \\
		H (2MASS)               	   & $12\fm 022$    \\
		K$_s$(2MASS)            	& $11\fm 958$    \\
		K$_p$(\textit{Kepler})  	& $13\fm 545$    \\
		E$_{B-V}$                    	   & $0\fm 117$    \\
		Period                 			       & 10.31\,d          \\
		$\pi$  (mas)            			    & 0.6606        \\
		\hline
	\end{tabular}
	\label{Table:inPar}
\end{table}
 
\section{Spectroscopic observations} \label{sec:spect}

We obtained spectra of KIC\,2306740 at 15 epochs with the
Intermediate dispersion Spectrograph and Imaging System (ISIS) on the
William Herschel Telescope (WHT), in July 2012.  These were timed to
provide the best possible coverage of the orbital phases, given that
the orbital period of 10.3\,d is significantly longer than the duration
of the observing run of 7 nights.  Two epochs occurred close to eclipse
when the velocity separation of the stars was small.  These measured
radial velocities are strongly affected by line blending and so were
not used in our analysis.

WHT equipped with the double-armed ISIS.
Spectra were taken simultaneously in the blue and red arms, covering
the regions around H$\gamma$ and H$\alpha$. In the blue arm we
used the grating H2400B, with a wavelength coverage of
4200 to 4550\,\AA. In the red arm the R1200R grating was used
and gave coverage of 6100 to 6730\,\AA.  The slit was set to 0.5\,arcsec
in order to limit the effects of telescope pointing errors so that a
resolving power of $R \approx 22\,000$ is achieved.  We used exposure
times of 1500\,s for all spectra to give a signal-to-noise (S/N) of
roughly 30\,per resolution element in the blue and 80 in the red.  We
bracketed each with spectra of CuAr+CuNe arc lamps for wavelength
calibration. 
The data were reduced using the {\sc pamela}  package \citep{Marsh1989}.

To measure the radial velocities (RVs) of the two stars from these
spectra we used standard cross-correlation (e.g.\ Tonry \& Davis 1979)
and its two-dimensional extension {\sc todcor}
\citep{1994ApJ...420..806Z}.  Synthetic template spectra were
calculated with the {\sc uclsyn} code (Smith 1992; Smalley et al.
2001) and {\sc atlas9} model atmospheres for metallicity, with a
$T_{\rm eff}$ of 5500\,K and no rotational broadening. For our final
RVs we adopt those given by standard cross-correlation.  These are very
similar to those obtained with {\sc todcor}.

\begin{table}
	\caption{Radial velocity measurements for KIC\,2306740.}
	\begin{tabular}{llcccc}
		\hline
		HJD         &   Phase   &   $V_1$  & $(O-C)_1$ &   $V_2$  &   $(O-C)_2$    \\
		(2456000+)  &           &   ${\rm km\,s^{-1}}$     & ${\rm km\,s^{-1}}$ &   ${\rm km\,s^{-1}}$  &   ${\rm km\,s^{-1}}$    \\
		\hline
		87.53965	&0.76260 &	71.76	&	1.98	&	-39.20	&	1.98	\\
		87.62467	&0.77085 &	71.16	&	2.91	&	-36.82	&	2.91	\\
		87.70422	&0.77857 &	67.24	&	0.46	&	-37.60	&	0.46	\\
		88.48349	&0.85417 &	51.19	&	0.08	&	-16.19	&	0.08	\\
		88.57612	&0.86316 &	48.36	&	-0.81	&	-15.99	&	-0.81	\\
		88.70547	&0.87571 &	45.49	&	-0.95	&	-12.52	&	-0.95	\\
		91.52893	&0.14965 &	-13.81	&	-1.63	&	55.10	&	-1.63	\\
		91.58456	&0.15504 &	-17.25	&	-4.00	&	56.01	&	-4.00	\\
		91.63997	&0.16042 &	-16.16	&	-1.85	&	60.03	&	-1.85	\\
		92.53903	&0.24765 &	-32.60	&	-2.67	&	77.83	&	-2.67	\\
		92.59339	&0.25292 &	-35.81	&	-5.08	&	77.17	&	-5.08	\\
		92.65527	&0.25893 &	-36.34	&	-4.73	&	78.21	&	-4.73	\\
		92.71038	&0.26427 &	-36.34	&	-3.97	&	79.82	&	-3.97	\\
		\hline
	\end{tabular}
	\label{Table:RV:data}
\end{table}

\begin{table}
	\caption{Spectroscopic orbital parameters of KIC\,2306740. The standard errors
		$\sigma$ are given in parentheses in the last digit quoted.}
	\begin{tabular}{llll}
		\hline
		Parameter                              & $e$ not fixed      & $e$ fixed at 0.301 \\
		\hline
		T$_0$/d                          	  	 & 2456399.19(24)   & 2456399.21(25)      \\
		P/d                              			  & 10.3069(78)        &10.3075(81)      \\
		e                               		 		& 0.322(22)          &0.301    \\
		$\omega$/rad                     	 & 4.81(3)            &4.80(3)      \\
		K$_1$/km\,s$^{-1}$                & 65.78(79)          &63.98(76)      \\
		K$_2$/km\,s$^{-1}$                & 72.84(87)          &70.86(84)      \\
		V$_o$/km\,s$^{-1}$                & 18.6(2)             &18.6(2)      \\
		q = ${m_1}/{m_2}$                  & 1.1073(90)          &1.1074(88)     \\
		a$_1 \sin i$/$\rm{R_{\odot}}$     	  & 12.682             &12.425       \\
		a$_2 \sin i$/$\rm{R_{\odot}}$     	  & 14.043             &13.761      \\
		m$_1 \sin^3i$/$\rm{M_{\odot}}$     & 1.304               &1.193      \\
		m$_2 \sin^3i$/$\rm{M_{\odot}}$     & 1.178               &1.077      \\
		\hline
	\end{tabular}
	\label{Table:RV:Orbit}
\end{table}

\begin{figure}
	\includegraphics[scale=0.2]{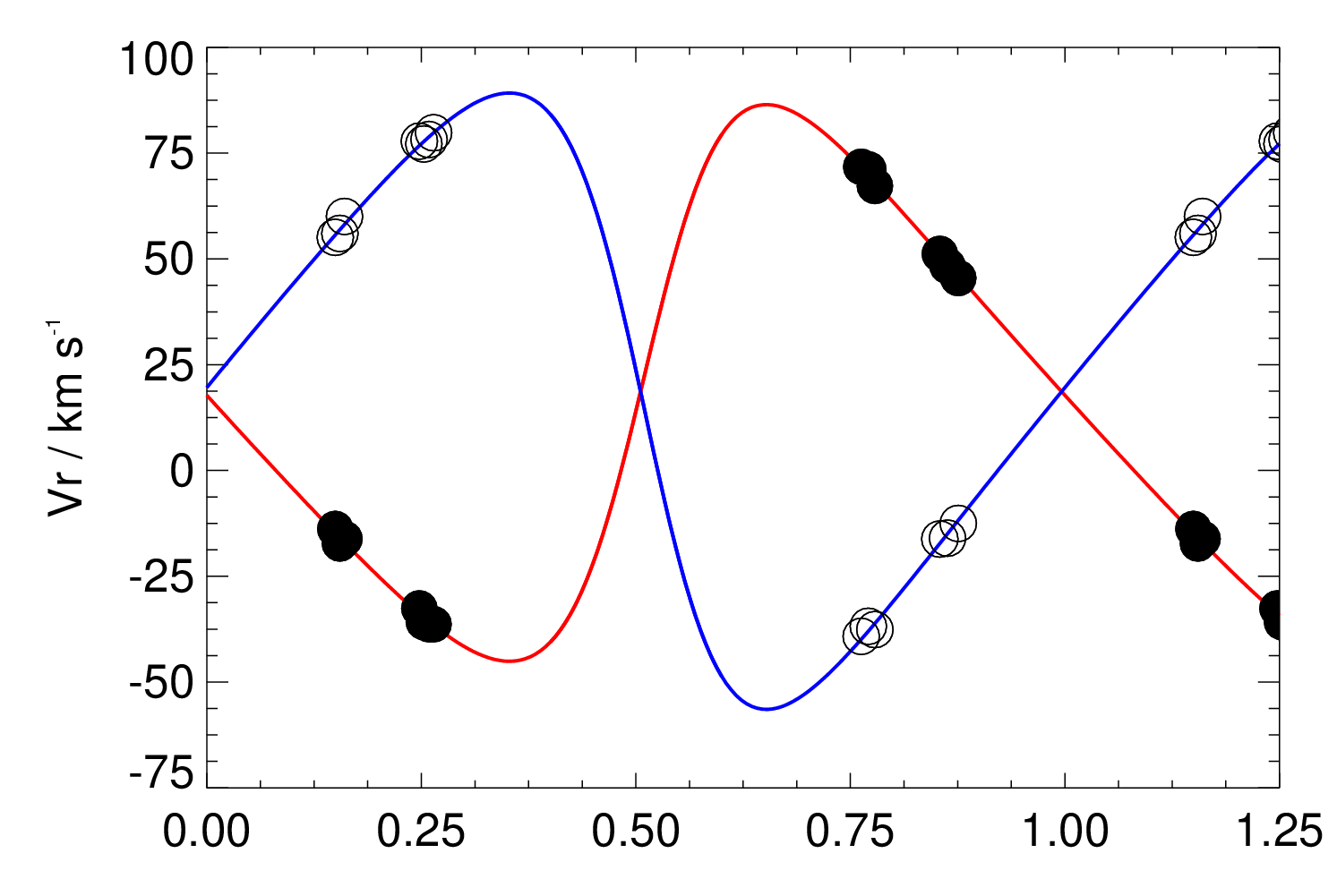}\\
	\vspace{-0.2cm}
	\includegraphics[scale=0.2]{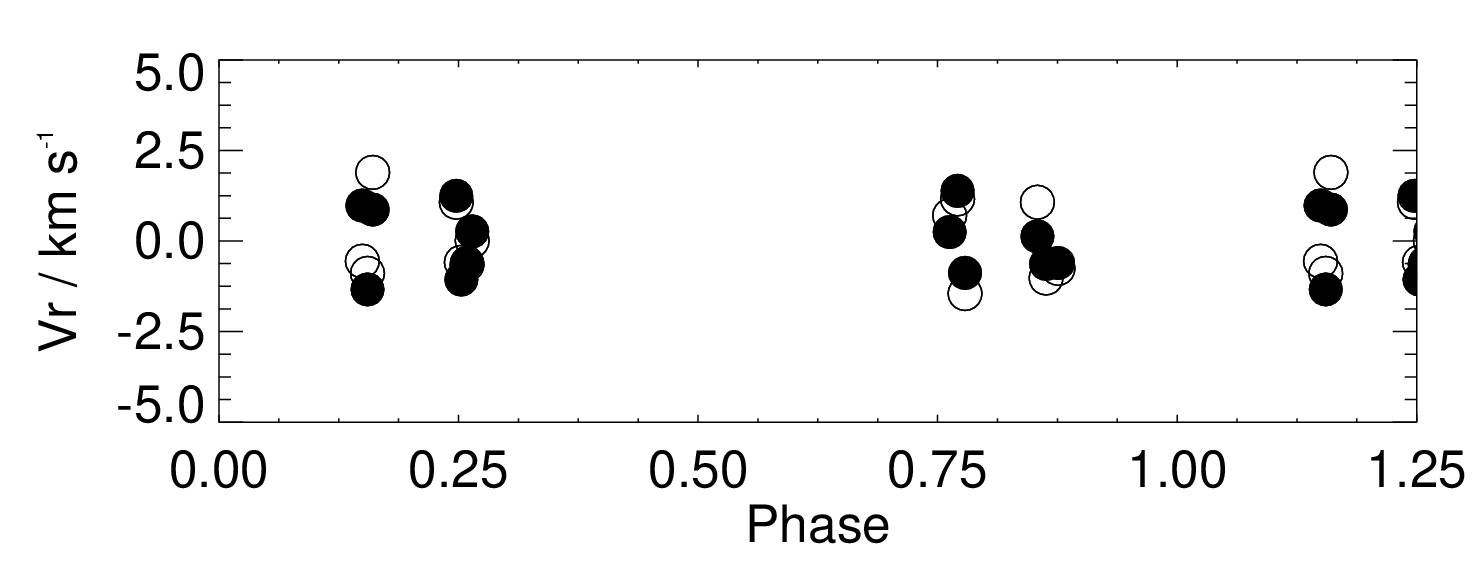}\\
	\caption{The radial velocity observations of KIC\,2306740 as a function of
		phase.  The filled and open circles represent the velocities of the
		primary and the secondary component, respectively.  The residuals are
		shown in the bottom panel.  The data are listed in
		Table \ref{Table:RV:data} and the curve fitting corresponds to the
		elements given in Table~\ref{Table:RV:Orbit}.}
	\label{Figure:RV}
\end{figure}

\section{\textit{Kepler} Observations of the system and Modeling of the light curve}\label{subsec:photometry}
\begin{figure}
    \begin{center}
	\includegraphics[scale=0.13]{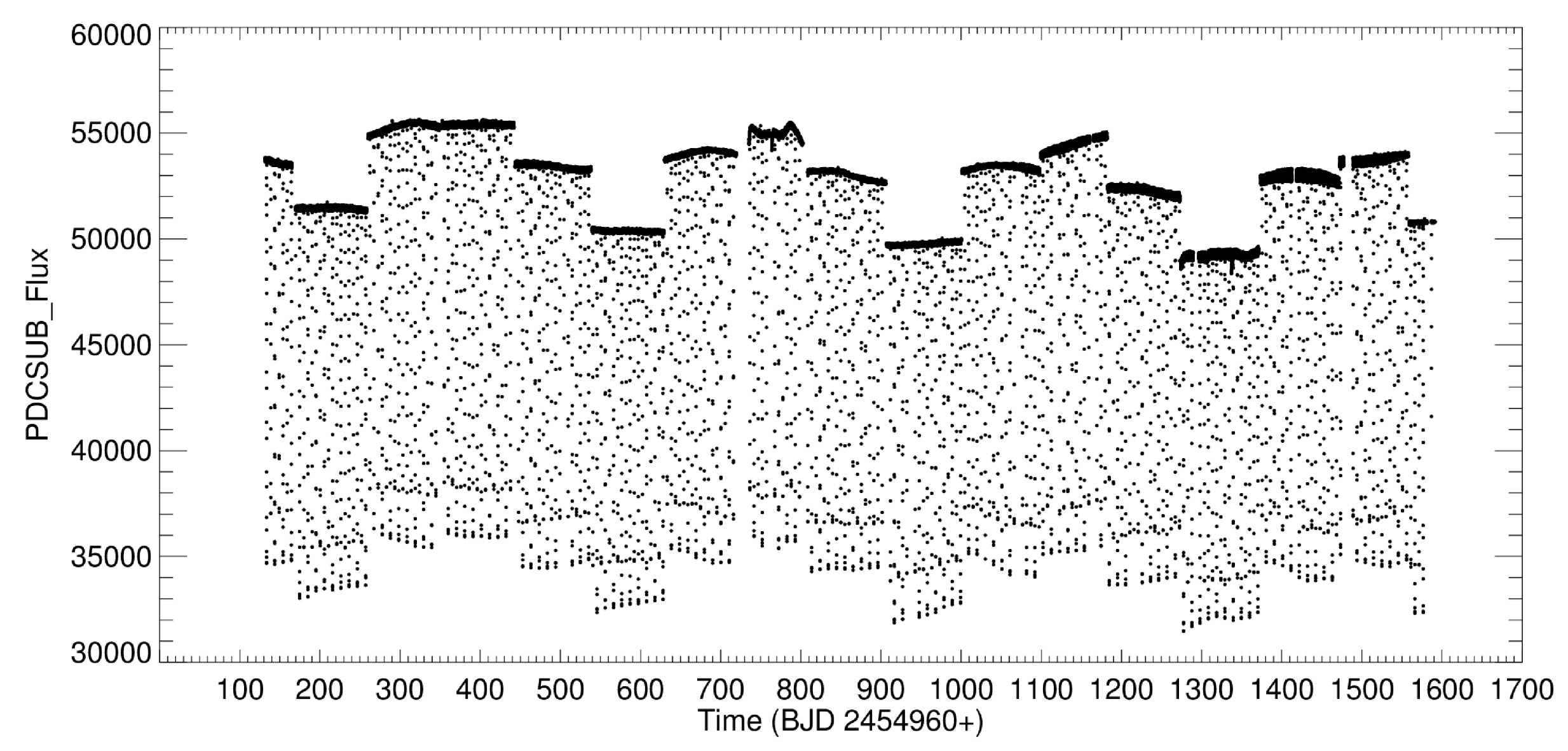} \\
	\includegraphics[scale=0.13]{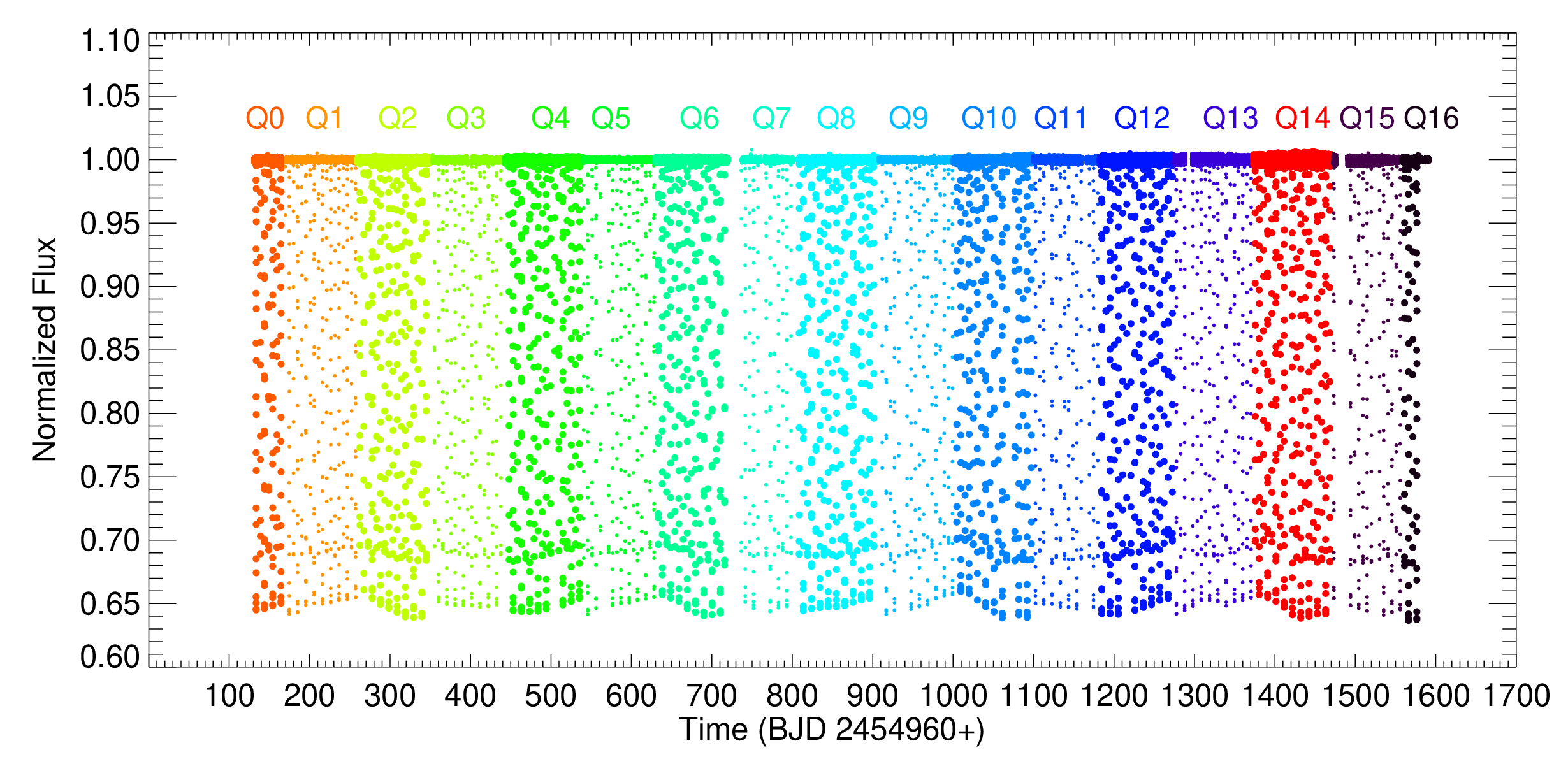} \\
	\caption{Kepler Q0 to Q16 observations of KIC\,2306740 raw data (upper panel) and de-trended (lower panel) data. The quarters are shown in different colors.}
	\label{Figure:Raw:Color}
	\end{center}
\end{figure}

The system was observed over approximately 1460\,days during seventeen quarters
(Q0 to Q16) with a long cadence (exposure time of about $30$\,min) and a
total of 64\,370 data points were obtained using the satellite. 
The light curve shows deep eclipses with periods of totality, plus periodic variations due to pulsations. 
{\it Kepler} satellite observations show some fluctuations due to common instrumental effects \citep{2010ApJ...713L..87J}. 
Using the techniques outlined in \citep{2010ApJ...713L..87J}, cotrending and detrending were applied to eliminate systematic variations.
We studied each quarter separately and, to de-trend the data, a third-order polynomial
fit was applied as we did in our earlier  {\it{Kepler}} study \citep{yakut2015,2019MNRAS.488.4520C}. 
The raw data of KIC\,2306740 is shown in Fig.~\ref{Figure:Raw:Color} (upper panel).  The de-trended normalized light variation is shown
in Fig.~\ref{Figure:Raw:Color} (lower panel).  The quarters are shown in different
colors.

\begin{figure}
	\includegraphics[scale=0.37]{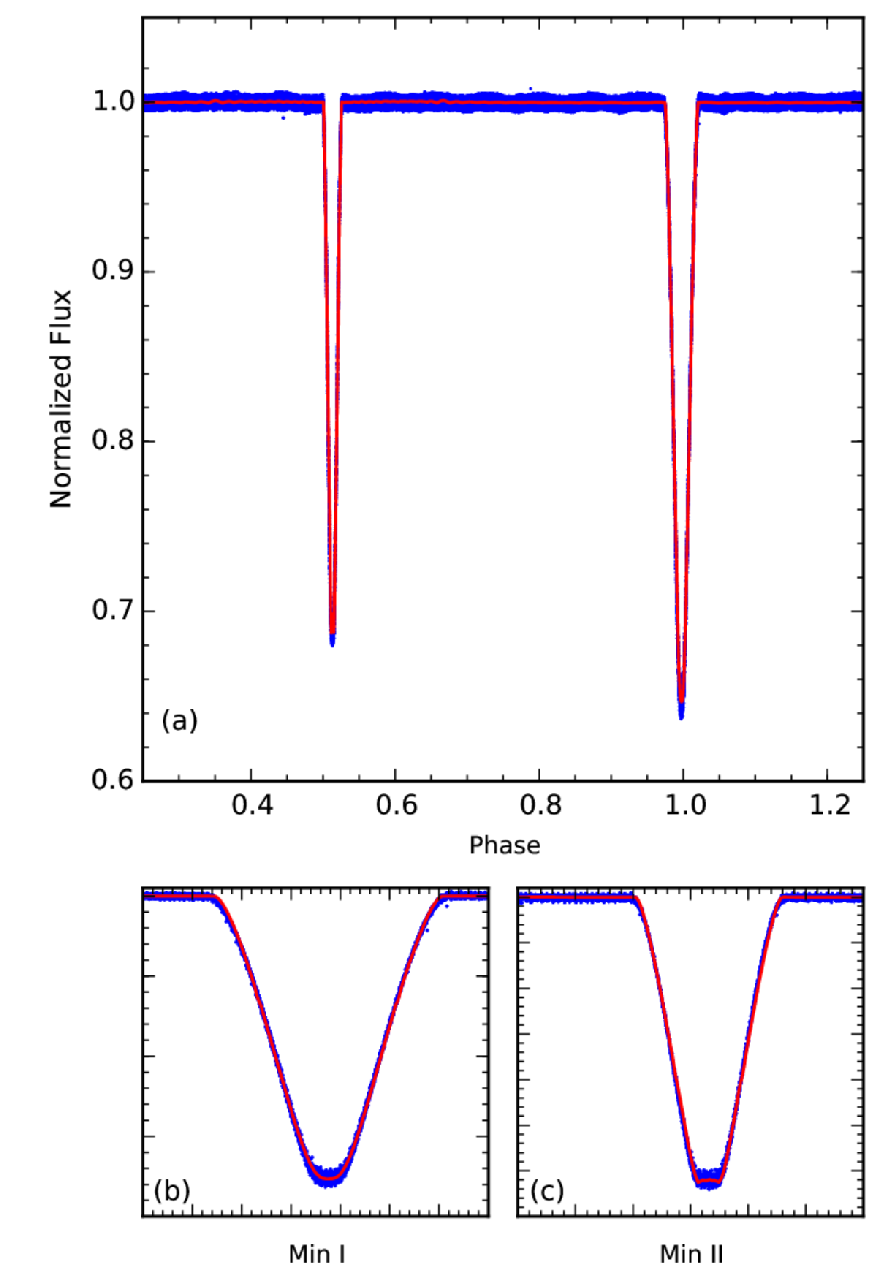} 
	\includegraphics[scale=0.14]{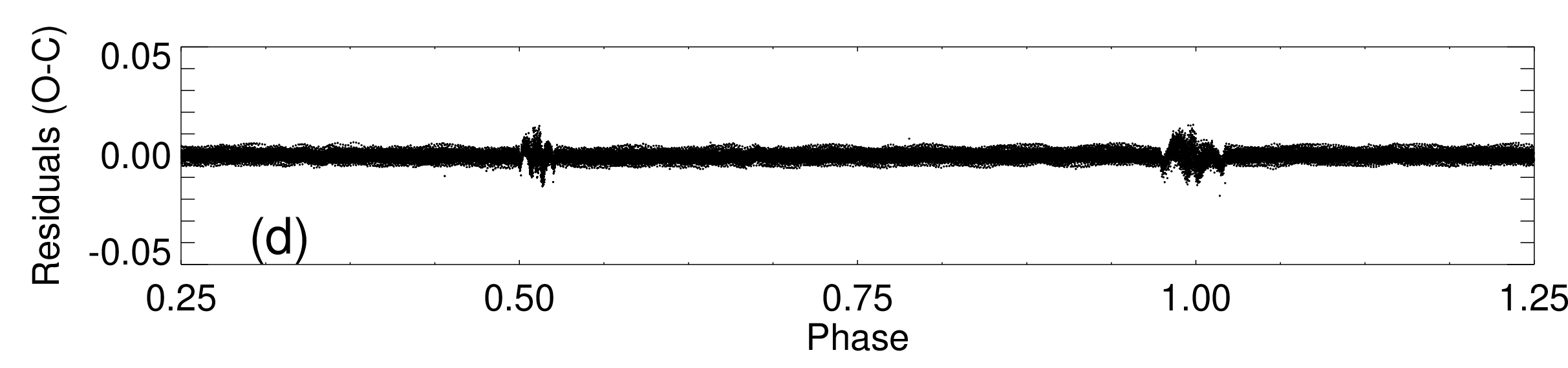} \\
	\caption{(a) \textit{Kepler} full-data set observation (blue dots) and computed (red line) light curve of
		the system.  Zoomed secondary~(b) and primary~(c) minima are shown to emphasize the agreement and residuals between the observed values and the corresponding LC model (d).} 
	\label{Figure:LC}
\end{figure}

Using the  \emph{Kepler} observations  we derived the linear ephemeris given in Eq.\ref{eq1}.
\begin{equation} \label{eq1}
\textrm{HJD~Min~I} = 24~56399^\textrm{d}.1227(2)+10^\textrm{d}.306988(2).
\end{equation}
During the calculation of the orbital phases in the Figures \ref{Figure:Raw:Color} - \ref{Figure:LC} and Table \ref{tab:lcfit} we used Eq.\ref{eq1}.

The {\it{Kepler}}  light curves along with the WHT RV curves were modeled simultaneously with the \textsc{jktebop} code\footnote{\texttt{http://www.astro.keele.ac.uk/jkt/codes/jktebop.html}} \citep[see][]{2004MNRAS.351.1277S, 2013A&A...557A.119S} 
and also with the \textsc{Phoebe} (Pr{\v s}a \& Zwitter 2005) program which uses the {\sc W--D} code \citep{wilson1971}. 
The curve dependent weights were assigned as described by \cite{1979ApJ...234.1054W}. 
We ran the code assuming a detached configuration. During analysis we iteratively solved the LC and
RV curves: the LC gave a more accurate estimate of the
eccentricity ($0.30 \pm 0.01$) than did the RV curves ($0.32 \pm 0.02$),
and so we fixed $e$ to this photometric solution when re-solving the
RV curves.  Even though {\it{Kepler}} data are very sensitive they are
all obtained in a single filter.  This prevents us from determining
accurate temperatures from multiple color analyses. 
Since the spectral data obtain is not sufficient to determine a precise temperature, the temperature of the hotter star was fixed to 6060 K found in \cite{Armstrong14}.

To determine the uncertainties of the measured properties of the KIC 2306740 system, we turned to \textsc{jktebop} as it is much faster than \textsc{phoebe}. \textsc{jktebop} and derivatives of the Wilson-Devinney code have been found to yield results consistent at the 0.2\% level or better in well-separated systems \citep{2020MNRAS.tmp.1795M}. For computational efficiency we phase-binned the data by sorting it into orbital phase and combining each group of 20 consecutive datapoints, giving a total of 3219 datapoints.

The phase-binned data and the radial velocities were modeled using \textsc{jktebop}. The fitted parameters were the fractional radii, orbital eccentricity, inclination, argument of periastron, central surface brightness ratio of the stars, third light, the linear limb darkening coefficient of the primary star, and the velocity amplitude of each star and systemic velocity of the system. Limb darkening was implemented using the quadratic law for both stars. Numerical integration was used to account for the fact that the data were obtained in long cadence by the \textit{Kepler} satellite \citep[see][]{2011MNRAS.417.2166S}. The fractional radii were fitted using their sum and ratio as these are less correlated. The orbital eccentricity $e$ and argument of periastron $\omega$, were fitted using the Poincar\'e parameters $e\cos\omega$ and $e\sin\omega$, for the same reason. Uncertainties were calculated for each fitted parameter and for each derived parameter in this study. 
This was done in two ways: using Monte Carlo and residual-permutation simulations \citep{2008MNRAS.386.1644S}. The uncertainties from the residual-permutation algorithm were found to be larger by typically a factor of 1.5 than those from the Monte Carlo algorithm, so were adopted as the final errorbars. 

Simultaneous LC and RV solutions were made using the full Q0--Q16 data and the analyses 
are summarized in Table~\ref{tab:lcfit}. In Fig.~\ref{Figure:LC} the computed light curves are shown by solid
lines.
\citet{prsa2011} gave preliminary orbital parameters for 1879 {\it{Kepler}} binary systems, including KIC\,2306740. They obtained a temperature ratio of 0.86, 
a sum of the fractional radii of 0.1374 and a $\sin\,i$ of 0.99919.
Our analysis includes RVs as well as much more extensive LCs, so the results given in Table~\ref{tab:lcfit} differ from those found by \citet{prsa2011}. 

\begin{table} \begin{center}
		\caption{\label{tab:lcfit} Fitted and parameters for KIC 2306740 from the \textsc{jktebop} analysis.}
		\begin{tabular}{l r@{\,$\pm$\,}l r@{\,$\pm$\,}l} \hline
			Parameter                                             &       \mc{Value}     \\
			\hline                                                
			\textit{Fitted parameters:} \\
			Sum of the fractional radii                           &    0.11102 & 0.00019 \\
			Ratio of the radii                                    &    0.7289  & 0.0025  \\ 
			Central surface brightness ratio                      &    0.9745  & 0.0057  \\
			Orbital inclination ($^\circ$)                        &    89.670  & 0.073   \\
			$e \cos \omega$                                       &    0.02329 & 0.00001 \\
			$e \sin \omega$                                       & $-$0.3002  & 0.0012  \\
			Third light                                           &    0.0918  & 0.0079  \\
			Velocity amplitude of cool component (km\,s$^{-1}$)     &    64.00   & 0.28    \\
			Velocity amplitude of hot component (km\,s$^{-1}$)   &    70.87   & 0.23    \\
			Systemtic velocity of cool component (km\,s$^{-1}$)     &    18.70   & 0.02    \\
			Systemtic velocity of hot component (km\,s$^{-1}$)   &    18.57   & 0.02    \\[3pt]
			\textit{Derived parameters:} \\
			Fractional radius of cool component                     &    0.06421 & 0.00006 \\
			Fractional radius of hot component                   &    0.04681 & 0.00017 \\
			Light ratio                                                  &    0.518  & 0.007  \\
			Orbital eccentricity $e$                              &    0.3011  & 0.0012  \\
			Argument of periastron $\omega$ ($^\circ$)            &  274.44    & 0.02    \\
			Mass ratio                                            &    0.903  & 0.004  \\
			\hline \end{tabular} \end{center} \end{table}

\section{Physical parameters of the components}  \label{sec:phypar}

\begin{table}
	\begin{center}
		\caption{Astrophysical parameters of KIC\,2306740. The standard
			$1\sigma$ errors of the last digits are given in parentheses.}
		\label{Table:PhysicalParameters}
		\begin{tabular}{lll}
			\hline
			Parameter                                                 	  & Cool Component      &Hot Component  \\
			\hline
			Mass ($M/M_{\odot}$)                         	   & 1.194(8)           & 1.078(7)      \\
			Radius ($R/{R_{\odot}}$)        					& 1.682(4)            & 1.226(5)      \\
			Temperature  $\log_{10} (T_{\rm eff}/K$)	& 3.764(18)            & 3.782(17)      \\
			Luminosity $\log_{10} (L/L_{\odot}$)                & 0.449(75)           & 0.260(72)      \\
			Surface gravity $\log_{10} (g/\rm{cm\,s^{-2}}$)  & 4.063(2)               & 4.294(4)      \\
			Bolometric magnitude (M$_b$)                               & 3.61(19)                & 4.08(18)          \\
			Absolute magnitude (M$_V$)                                   & 3.69                & 4.09         \\
			Semi-major axis ($a/{R_{\odot}}$)        &~~~~~~~~~26.201(44) &        \\
			\hline
		\end{tabular}
	\end{center}
\end{table}

The physical parameters of a binary system can best be derived if it
is a double-lined eclipsing binary system with an accurate light
curve. Hence, the detached binary system KIC\,2306740, for which the 
photometric and spectroscopic data are both of high precision, is
excellent for accurate determination of its parameters.  The detailed
LC solution of the system indicates that the
stars are well detached from their Roche lobes (see Section~3).  

In this study, RVs and LCs were analysed
simultaneously and the orbital parameters of the system were obtained.
With the measurements given in Tables~\ref{Table:RV:Orbit}
and ~\ref{tab:lcfit} we can estimate the physical parameters of
the components given in Table ~\ref{Table:PhysicalParameters}. 
We have used the \textsc{jktabsdim} code \footnote[1]{jktabsdim: www.astro.keele.ac.uk/~jkt/codes/jktabsdim.html} to estimate the physical parameters of the components, with uncertainties propagated from the LC and RV solution using a perturbation analysis. The nominal physical constants and solar properties recommended by the IAU were used \citep{2016AJ....152...41P}. All the calculated parameters of the binary system are summarized in Table~\ref{Table:PhysicalParameters} with their estimated errors. 

The masses of the two stars are slightly greater than solar and the radii are significantly larger. The hotter star is the less massive and has the smaller radius. This indicates that the larger star is near the end of its main-sequence lifetime. We discuss this in Section~6.

\section{Light variation outside eclipse} \label{sec:lcmax}

\begin{table}
	\begin{center}
		\scriptsize
		\caption{Computed genuine frequencies, amplitudes and phase shifts of the solution.
			Frequencies with signal-to-noise ratios ($S/N$) exceeding~4 are considered as significant.}
		\label{Table:Freq}
		\begin{tabular}{llllllll}
			\hline
			Frequency & & Amplitude   &   & Phase     &   & S/N  \\
			/d$^{-1}$ & & /mmag      &   & $\phi$    &   &                 \\
			\hline
			1.36410248	 &	(1)	 &	2.9497	&	(1)	&	3.6848	&	(1)	&	860		\\
			1.36287801   &	(1)	  &	1.0037	  &	(1)	  &	 1.5393	&	(3)	&	224		\\
			1.36930353	&  (2)	&  0.3028	&	(1)	&	2.9146	&	(8)	&	82		\\
			0.29110156	 &	(4)	 &	0.1311	  &	(1)	   &  5.7778	&	(19)&	32		\\
			0.0970908	 &	(5)	 &	0.1103	 &	(1)	  &	 1.9516	&	(25)&	22		\\
			\hline
		\end{tabular}
	\end{center}
\end{table}

Investigating of the sinusoidal brightness variation in the LC requires the extraction of the effects of binarity from the LC.
After making the simultaneous LC and RV analysis, 
we subtracted the binary model from the observations.
The residuals from the phased light variation of the binary are plotted in the Fig.~\ref{Figure:Puls}.
The oscillatory pattern can be seen clearly.

\begin{figure}
	\includegraphics[width=8cm,height=6cm]{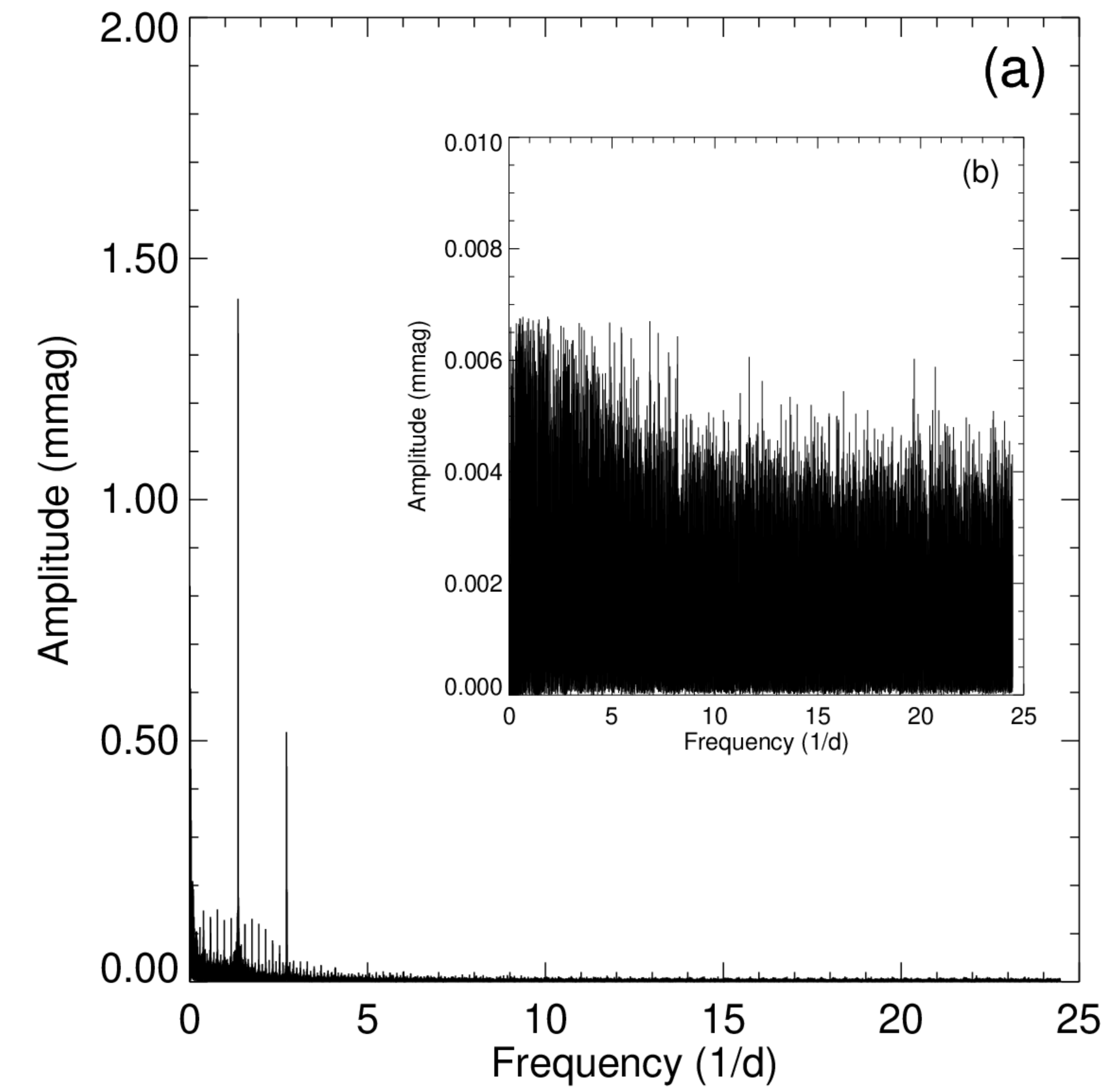}\\
	\caption{Amplitude A spectrum of the components before~(a) and after~(b)
		pre-whitening of all frequencies $\nu$.}
	\label{Figure:Amplitude}
\end{figure}

\begin{figure}
	\includegraphics[scale=0.2]{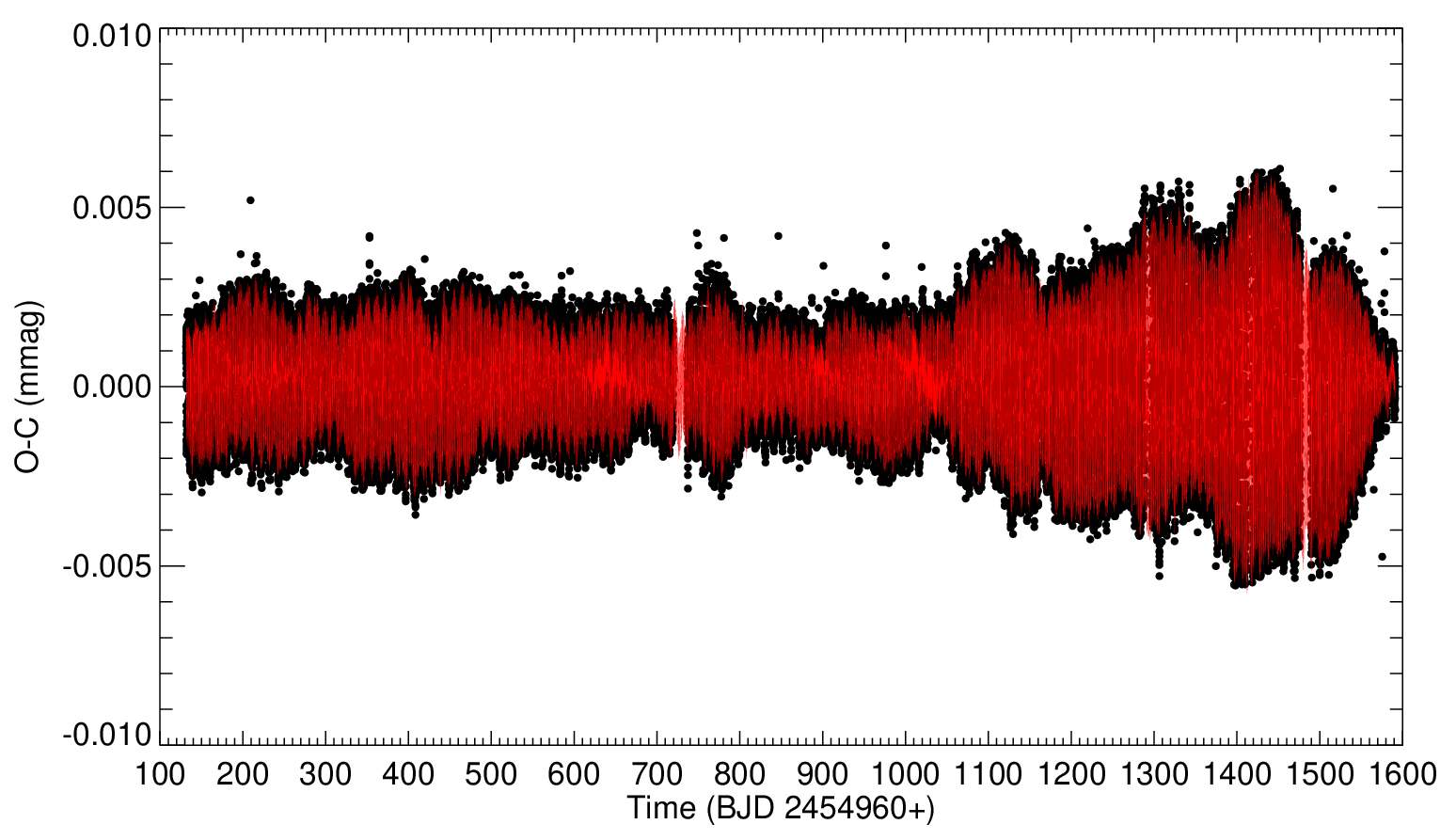}\\
	\includegraphics[scale=0.35]{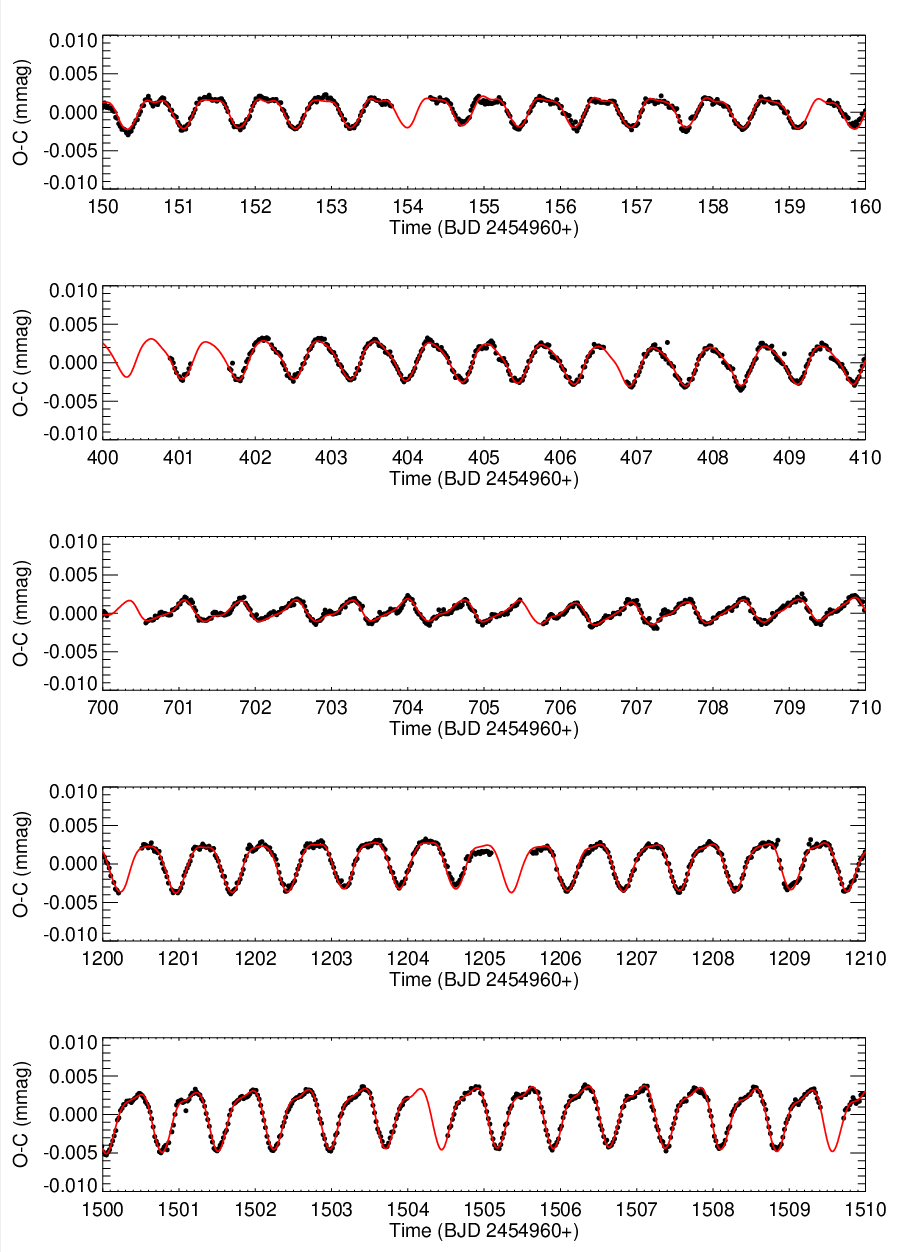}\\
	\caption{Part of data used in the frequency analysis (upper panel) is zoomed for different time intervals (lower panel).}
	\label{Figure:Puls}
\end{figure}

Since programs using modern light curve modeling were not perfect at representing \textit{Kepler} data, we performed a frequency analysis on all the long cadence data obtained out-of-eclipse using the {\sc SigSpec} \citep{2007A&A...467.1353R} and {\sc{Period04}} \citep{LenzBreger05} codes, which are based on classical Fourier analysis.
A signal-to-noise ratio $S/N > 4$ threshold was chosen as a criterion
to consider a frequency as significant \citep{Breger11}.  We searched for
significant peaks in the frequency interval from~0 to the Nyquist
frequency of $25\,\rm d^{-1}$ but found no meaningful peak above
$4\,\rm d^{-1}$. Fig.~\ref{Figure:Amplitude}a shows the amplitude spectrum before
pre-whitening of any frequency between~0 and~$25\,\rm d^{-1}$.
Higher amplitude peaks gather below a frequency of $5\,\rm d^{-1}$.
We continued to obtain pre-whitened frequencies until the signal
amplitudes fell below four times the average noise.
Fig.~\ref{Figure:Amplitude}b represents the spectra after
pre--whitening.

Table~\ref{Table:Freq} lists these genuine
frequencies, their amplitudes, phases and S/Ns sorted by decreasing
amplitude. Signal-to-noise ratios were computed over an interval of $5\,\rm d^{-1}$.  In the top
panel of Fig.~\ref{Figure:Puls} the agreement between 100~calculated
frequencies and the observational data is plotted for almost 4\,yr.
The bottom panel illustrates the zoomed data of 10\,d for clarity. 
The analysis resulted in the detection of five genuine and more than
100 combination frequencies.

What could be the mechanism that caused a change in the maximum amplitude of the KIC 2306740 system? Generally, such changes may result from stellar pulsating and/or periodic changes. In binary star systems, inhomogeneous structures (e.g. stellar spots) on the surface of one or both of the component stars can cause changes in the light curve in asynchronous situations, known as rotational variability. For the frequencies obtained from the Fourier analysis of the light variation of the KIC~2306740 system (Table 6), periods of approximately 0.7\,d, 3.4\,d and 10.3\,d were obtained. The 10.3\,d period is related to the orbital period and the 3.4\,d period is related to the spin period of the stars (see Section \ref{sec:results}). The source of the 0.7\,d periodicity may be $\gamma$\,Dor-type pulsation or spot modulations on one or both of the component(s).
Besides, looking into out-of-eclipse of light variation of the system we analyzed minima phases of the light curves. There is a variation with an amplitude of 0.03 at the primary minimum and a variation with an amplitude of 0.013 at the secondary minima. However, residual data is not sufficient to estimate new frequencies. 

\section{Results and Conclusion} \label{sec:results}

We have modeled the light and radial velocity curves of the
well-detached binary system KIC\,2306740 and determined its orbital
and physical parameters (Tables~\ref{tab:lcfit} and \ref{Table:PhysicalParameters}).
Solutions indicate that the more massive and cooler component of the
system is more evolved than the hotter less massive component.  
Recently, Gaia gave a parallax of 0.6606$\pm$0.0165\,mas \citep{2018yCat.1345....0G} for the KIC 2306740 (Gaia DR2 2051885033280089216). Using this Gaia parallax and our results we derive a distance modulus of $m_V-M_V=11\fm28$ and magnitudes for the components and the system of $V_1=14\fm97\pm0.10$, $V_2=15\fm50\pm0.12$ and $V_{\rm total}=14\fm50\pm0.11$.
The new stellar parameters and reddening, reveal the distance of the system to be 1.53 kpc, which is very close to the distance obtained by Gaia as 1.51 kpc.
With these results we can add the system to the list of well-determined binary
stars. The maximum light shows cyclical variations.

\begin{figure}
	\includegraphics[width=7cm,height=6.15cm]{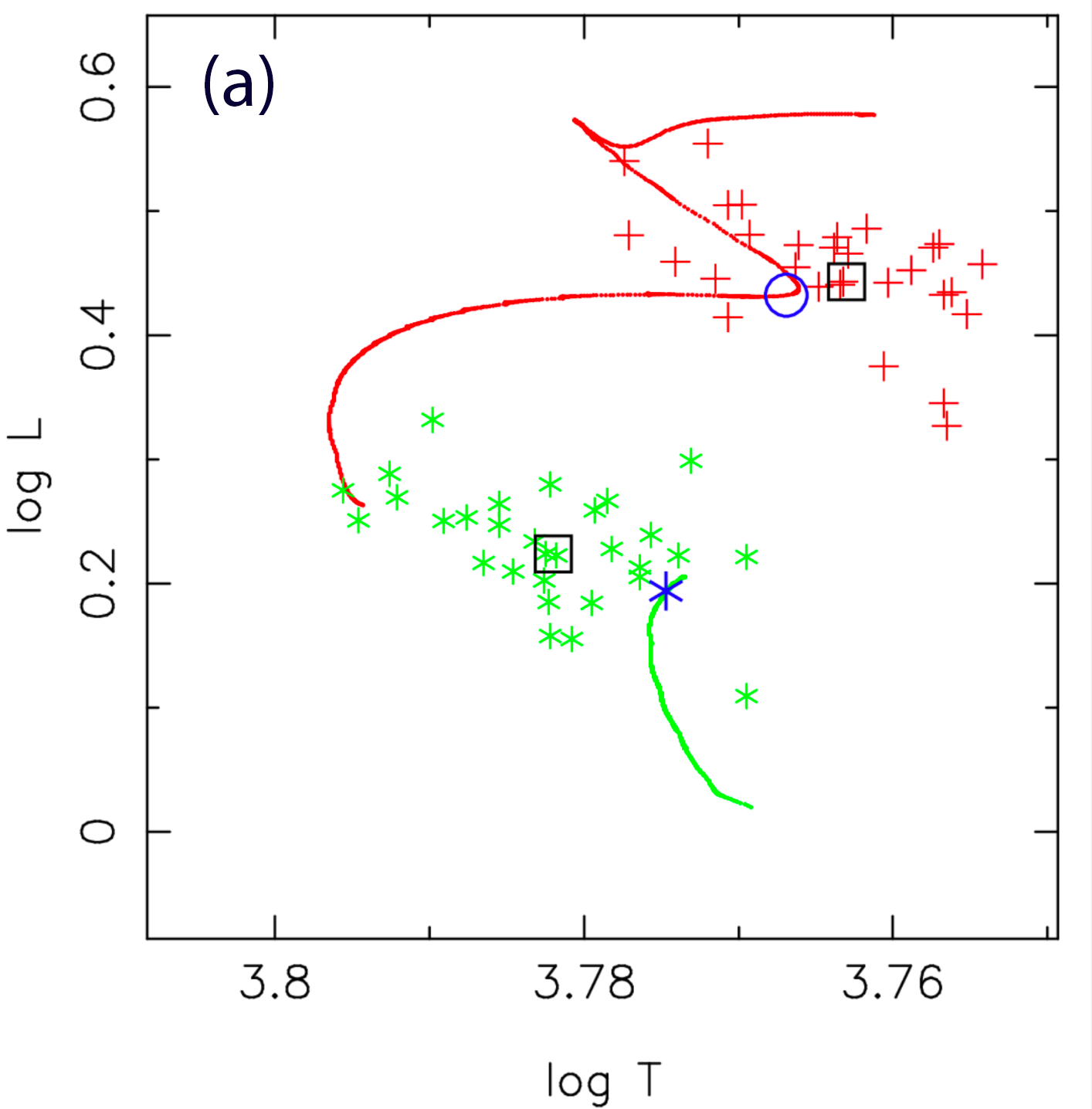}\\
	\includegraphics[width=7cm,height=6.15cm]{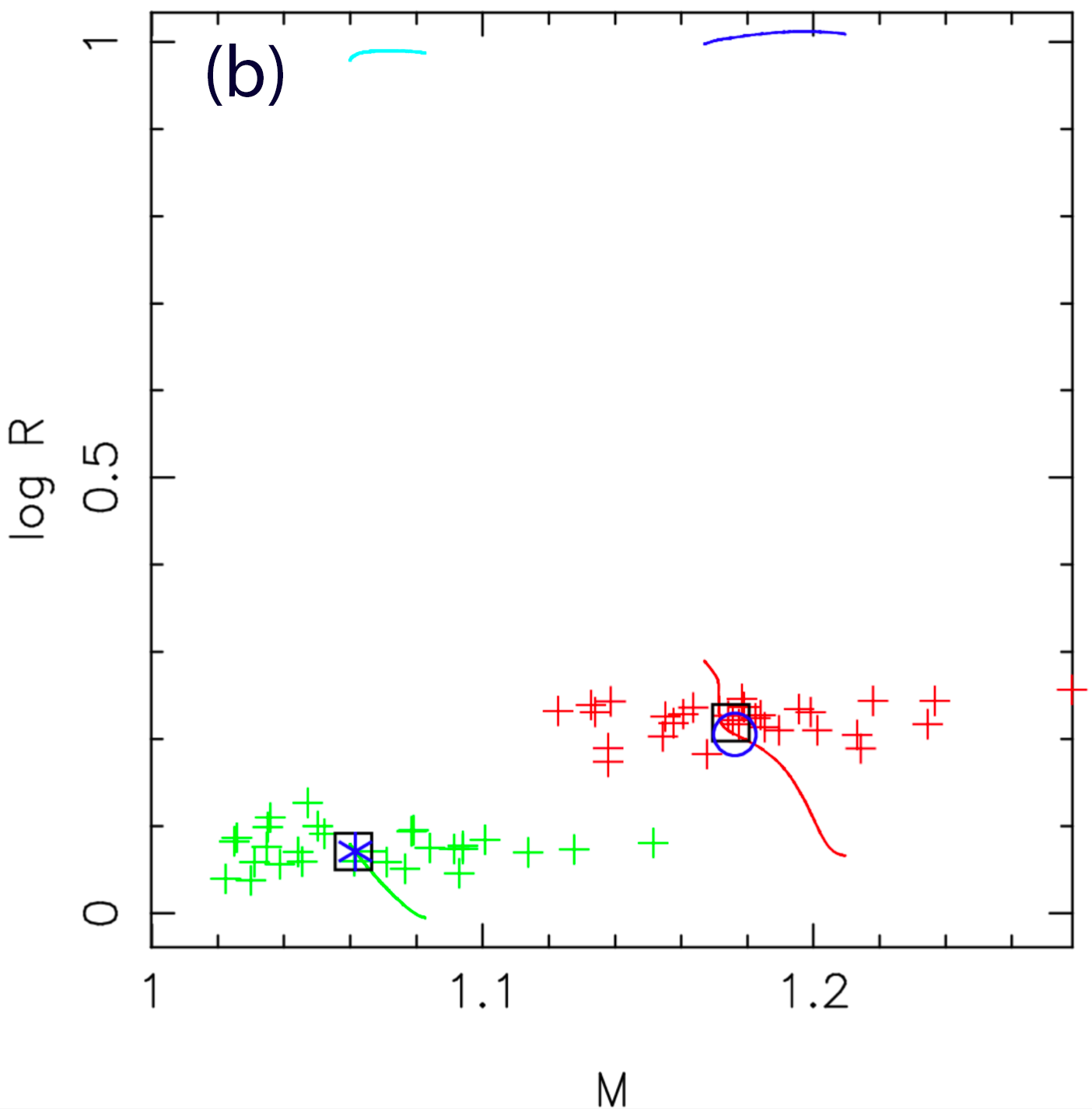}\\
	\includegraphics[width=7cm,height=6.15cm]{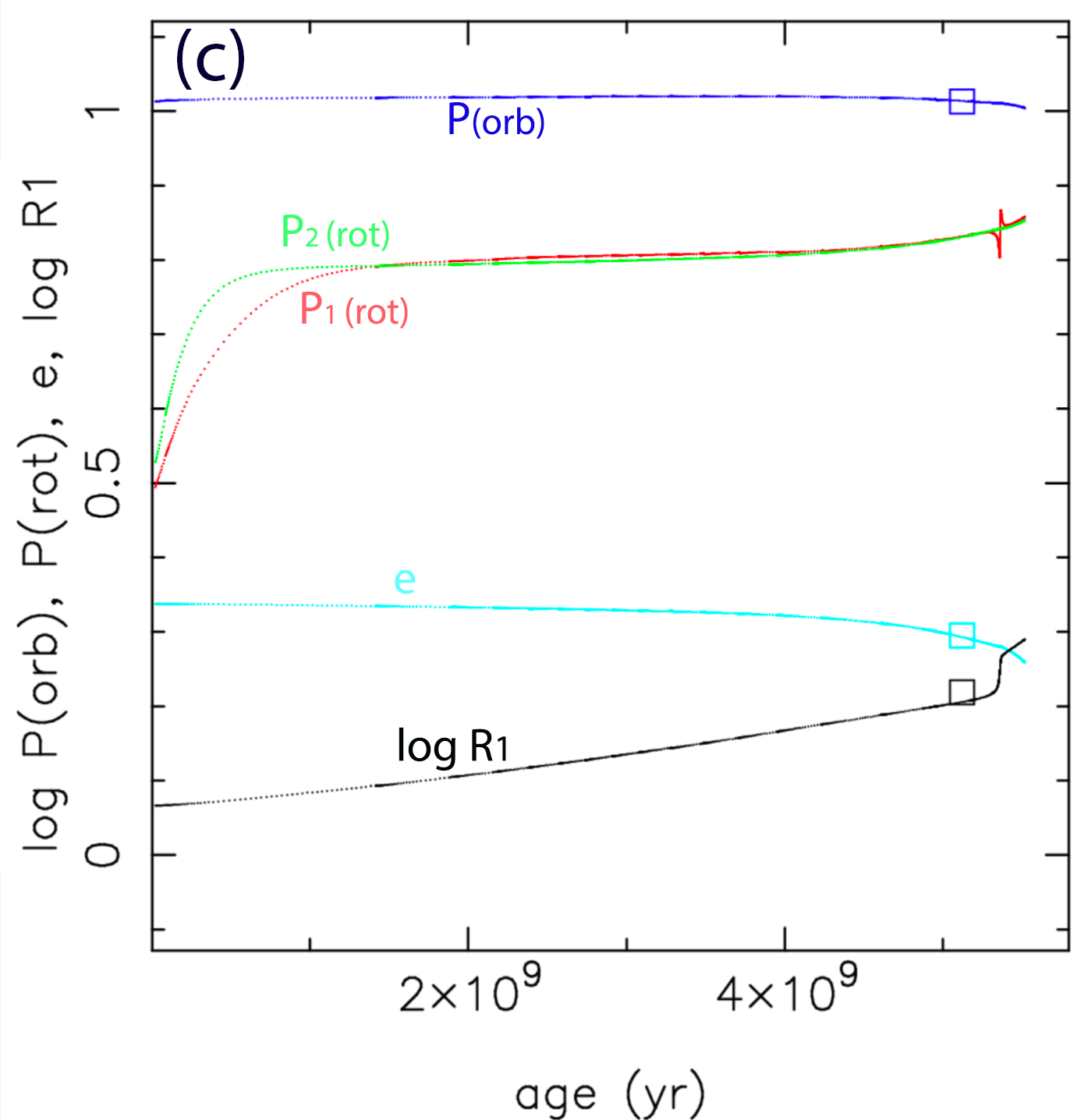}
	\caption{Evolutionary tracks for our best model of KIC\,2306740. Our data for KIC\,2306740 are shown as squares.  
	Panel (a) shows the  evolution in the $\log L - \log T$ plane.  The primary's and secondary's tracks are shown with red and green lines, respectively.  
	The extent of the red and green pluses represent roughly the observational uncertainties.  
	Panel (b) shows the primary's radii (red) and lobe radii (dark blue) and secondary's radii (green) and lobe-radii (light blue) as a function of mass. 
	Panel (c) shows the time-evolution of orbital period (dark blue), spin period of the primary (red) and the secondary (green), eccentricity (light blue) and radius of primary (black).}
	\label{Figure:EV}
\end{figure}

Out-of-eclipse light variations have been obtained after
eliminating the effects of binarity for the established orbit.
A frequency analysis of these data revealed more than 100~frequencies of
which~5 are significant non-combination frequencies. 
The results from this study indicate that the system KIC\,2306740 
may contains a non-radial pulsating  $\gamma$ Dor type star which pulsates in high-order gravity-modes.
Considering the other frequencies ($\sim 3.4$ d, half of the spin periods) obtained, we have found that this is related to the predicted spin period of $\sim 6.8$ days (see Fig.~\ref{Figure:EV}).

we have found that the component stars are related to the spin periods ($\sim 6.8$ d) obtained (see Fig. 6c). 
This tells us that at least one of the components has changed due to inhomogeneous structures on its surface. 
This also tells us that spot modulation is a possible explanation for the variation at the maxima phases, instead of pulsations.
Using the parameters we have obtained ( Table \ref{Table:PhysicalParameters}) show that the components of binary system KIC 2306740 is outside or on the edge of the instability zone in the HR diagram. Therefore, the light variation seen in this system is likely caused by spot modulation.

We use this system to test the theory of modeling stellar interiors
by comparing with its the observational properties. Models were
constructed with the  {\sc ev} code \citep{eggleton2002} and its much more
powerful  {\sc twin} variant \citep{yakut2005,eggleton2006,eggleton2010},
both of which are based on the Cambridge STARS code \citep{eggleton1971,eggleton1972,eggleton1973,pols1995}.
In single-star evolution, as in the STARS code, the effects of rotation and magnetic dynamo
activity on mass loss and the proximity of the companion are usually
not considered.  {\sc twin} allows various non-conservative processes to be
applied to the primary component of a binary system.  {\it Both}
components are modeled {\it simultaneously} so that the effects of
tidal friction, magnetic dynamo activity, and hence mass loss, can be
incuded according to a self-consistent prescription.  Mass loss
carries off angular momentum by way of magnetic braking.

In the case of non-conservative evolution it is hard to find
initial parameters that would lead to the current binary system \citep[see][]{EY2017}.
However we can reasonably assume that the initial masses were larger
than now.  Table~\ref{Table:An Evolutionary Model} gives parameters for a model that is slightly
metal-rich ($Z=0.03$) compared to the Sun.  This metallicity gives a
somewhat better fit than solar.

After some experimentation, we evolved a pair of stars with initial masses
of 1.21 and 1.08\,$\rm M_{\odot}$, each with spin periods of 3.0\,d, an
eccentricity of 0.34 and an orbital period of 10.30\,d.  We expect the
orbital period to decrease as the orbit circularizes and as angular
momentum is lost by wind mass loss and magnetic braking.  However the
period also {\it increases} on account of the orbit acquiring some
angular momentum from the spins of the stars by tides. The particular
model of the non-conservative processes in the {\sc twin} code led the
system to evolve to roughly the observed masses and radii in about
$5.1\,$Gyr.

\begin{table}
	\begin{center}
		\caption{An Evolutionary Model for KIC 2306740.}
		\label{Table:An Evolutionary Model}
		\begin{tabular}{lrll}
			\hline
			Parameter	                		  &	Zero age  	 	&	Age 5.12 Gyr    	 &	Observed \\
			\hline
			$P/\rm d$	               			  &\	10.30	     	&	10.32	                  &	10.307	 \\
			$e$	                        		      &\	0.337	    	&	0.293	                  &	0.301	\\
			$M_1/M_\odot$ 	            	&\	1.210	     		&	1.176	                   &	1.194	\\
			$\log_{10}(R_1/R_\odot)$   &\	0.058	    	  &	0.206	                    &	0.225	\\
			$\log_{10}(L_1/L_\odot)$	&\	0.208	    	   &	0.432	                 &	0.449	\\
			$\log_{10}(T_1/\rm K)$	     &\	3.785	     		&	 3.767	                   &	3.764	\\
			$M_2/M_\odot$ 	         	   &\	1.083	   		   &   1.062	                  &  1.078    \\
			$\log_{10}(R_2/R_\odot)$  & --0.013	      		 &	  0.071	                    &	0.088	\\
			$\log_{10}(L_2/L_\odot)$   &--0.005	      		 &	  0.194	                    &	0.260	\\
			$\log_{10}(T_2/\rm K)$	    &\	3.767	     		&	3.775	                  &	3.782	\\
			\hline
		\end{tabular}
	\end{center}
\end{table}

Fig.~\ref{Figure:EV} shows evolutionary tracks in three planes, (a) the $\log L -
\log T$ plane or theoretical HR diagram, (b) the $\log R - \log M$
plane and (c) the $\log P - \rm age$ plane for both spin and orbital
periods periods together with eccentricity $e$ and $\log R_1$.  Our
data for KIC\,2306740 are represented by squares in panels (a)
and (b).  Each square is surrounded by a cloud of plusses generated by
a Gaussian random number generator to illustrate the extent of the
standard errors in the basic observational data ($K_1$, $K_2$, $T_1$,
\dots). The evolutionary tracks for the 
more massive star (*1) are red and for less massive star (*2) green.
A blue circle approximately marks the best fit to
$*1$ and a blue asterisk the coeval point for $*2$.  Panel (b) shows
the Roche lobe radii as well as the actual radii.  Neither star is within a
factor of 5 of its Roche radius.  Panel (c) shows the spin
periods in red and green.  Both were started arbitrarily at $3\,$d and
rather rapidly evolved to about $6\,$d before reaching a plateau.
During this fairly rapid initial spin-down the components lost about
$0.03$ and $0.02\,M_{\odot}$ of their masses.  The logarithm of the orbital period is
dark blue, the eccentrcity is light blue and the radius of star 1 is black.
Table~\ref{Table:An Evolutionary Model} shows the evolutionary changes in some major variables.  Our
overall conclusion from Fig.~\ref{Figure:EV} is that the fit of to the theoretical
model is acceptable at a 1$\sigma$ level but it would be better if the model temperatures matched more closely those observed.

\acknowledgments
We are very grateful to an anonymous referee for comments and
helpful, constructive suggestions, which helped us to improve
the paper. The authors gratefully acknowledge the numerous people who have helped
the NASA {\it{Kepler}} mission possible. This study was supported by
the Turkish Scientific and Research Council (T\"UB\.ITAK 117F188).  
DK is grateful to the Astronomy Department of the University of Geneva (Geneva Observatory) for the kind hospitality 
during her visit and gratefully acknowledge the support provided by the T\"UB\.ITAK-B\.IDEB 2211-C and 2214-A scholarships. 
CAT thanks Churchill College for his Fellowship. 
KY would like to acknowledge the contribution of  COST (European Cooperation in Science and Technology)  Action CA15117 and CA16104. 


%

\vspace{5mm}
\facilities{Kepler, William Herschel Telescope (WHT)}


\software{{\sc todcor}  \citep{1994ApJ...420..806Z} , {\sc pamela}  \citep{Marsh1989}, 
	{\sc SigSpec} \citep{2007A&A...467.1353R}, {\sc{period04}} \citep{LenzBreger05}, 
	 {\sc twin} \citep{yakut2005,eggleton2006,eggleton2010},
 {\sc Cambridge STARS Code}  \citep{eggleton1971,eggleton1972,eggleton1973,pols1995}
}



\end{document}